\documentclass[aps,final,notitlepage,oneside,onecolumn,nobibnotes,nofootinbib,superscriptaddress,
noshowpacs,centertags,showpacs,showkeys]{revtex4-1}

\usepackage[english]{babel}
\usepackage{latexsym}
\usepackage{amssymb}
\usepackage{amsmath}

\begin{document}

\title{Conformal gravity and ``gravitational bubbles''}

\author{V.\;A.\;Berezin}\thanks{e-mail: berezin@inr.ac.ru}
\affiliation{Institute for Nuclear Research of the Russian Academy of Sciences
60th October Anniversary Prospect 7a, 117312 Moscow, Russia}
\author{V.\;I.\;Dokuchaev}\thanks{e-mail: dokuchaev@inr.ac.ru}
\affiliation{Institute for Nuclear Research of the Russian Academy of Sciences
60th October Anniversary Prospect 7a, 117312 Moscow, Russia}
\affiliation{National Research Nuclear University MEPhI (Moscow
Engineering Physics Institute), 115409 Moscow, Russia}
\author{Yu.\;N.\;Eroshenko}\thanks{e-mail: eroshenko@inr.ac.ru}
\affiliation{Institute for Nuclear Research of the Russian Academy of Sciences
60th October Anniversary Prospect 7a, 117312 Moscow, Russia}

\date{\today}

\begin{abstract}
We describe the general structure of the spherically symmetric solutions in the Weyl conformal gravity. The corresponding Bach equations are derived for the special type of metrics, which can be considered as the representative of the general class. The complete set of the pure vacuum solutions, consisting of two classes, is found. The first one contains the solutions with constant two-dimensional curvature scalar, and the representatives are the famous Robertson--Walker metrics. We called one of them the ``gravitational bubbles'', which is compact and with zero Weyl tensor. These ``gravitational bubbles'' are the pure vacuum curved space-times (without any material sources, including the cosmological constant), which are absolutely impossible in General Relativity. This phenomenon makes it easier to create the universe from ``nothing''. The second class consists of the solutions with varying curvature scalar. We found its representative as the one-parameter family, which can be conformally covered by the thee-parameter Mannheim--Kazanas solution. We describe the general structure of the ene\-rgy-mo\-men\-tum tensor in the spherical conformal gravity and construct the vectorial equation that reveals clearly some features of non-vacuum solutions.
\end{abstract}
\maketitle

\section{Introduction}

The conformal gravity was invented by H.\;Weyl in 1918 \cite{Weyl}. His motivation was to combine the gravitational and electromagnetic fields into one unified theory. Then such a theory was rejected by A.\;Einstein and H.\;Weyl because it was recognized that the conformal invariance allows only for massless particles to exist. But, nowadays, this obstacle can be overcome by means of the Higgs mechanism for generating particle masses \cite{tHooft14}. The conformal gravity came again into play in 1989 due to the paper by P.\;D.\;Mannheim and D.\;Kazanas \cite{ManKaz89}, who obtained the static spherically symmetric vacuum solution, depending on three arbitrary constants and generalizing the Schwarzschild solution in General relativity (see also \cite{Edery98,Verbin09}). What was really new is the appearance of the term linear in radius, what forced to remember the Mach's principle.

We decided to study the conformal gravity attentively after becoming aware (quite accidentally) that all the homogeneous and isotropic space-times (i.e., described by the Robertson-Walker line element) have zero Weyl tensor. This means that they are solutions to the vacuum conformal gravity field equations. Such an empty high symmetric space-time is a good candidate for the creation of the universe ``from nothing'' (the very possibility of which was first proposed by A.\;Vilenkin~\cite{Vil82}). The idea that the initial state of the universe should be conformal invariant, is advocated also by R.~Penrose \cite{Penr10,Penr14}, and G.`t Hooft \cite{tHooft14}. But, this empty universe should be filled with the particles (necessarily massless prior to the spontaneous breaking of the conformal symmetry). How? - Due to the quantum fluctuations that will cause the local deviation of the Weyl tensor from zero. This, in turn, will cause the creation of massless particles (as wa shown first by Ya.\;B.\;Zel'dovich, I.\;D.\;Novikov and A.\;A.\;Starobinsky \cite{ZeldNovStar1974}) which, moving with the speed of light, will deviate the Weyl tensor form zero in much more large region, and so on, and so forth. For how long this could last, and what could be the result? As was shown by Ya.\;B.\;Zel'dovich~\cite{Zeld70} (See  also \cite{ZeldStar71,LukashStar74}), the process of particle creation can make the universe isotropic. The isotropization and homogenization will lead to decreasing of both the creation rate and the square of the Weyl tensor. But, in conformal gravity such solution are the vacuum ones! This controversy can be removed by taking in account not only the energy-momentum tensor of the already created particles, but also the quantum corrections coming from the vacuum polarization and trace anomaly \cite{ZeldStar71,Parker69,GribMam69,ZeldPit71}. In one-loop approximation the corresponding counter-terms can be incorporated into the action integral as the curvature scalar (manifesting the renormalization/emergence of Newton's constant), the cosmological constant and the term, quadratic in Ricci tensor. This was foreseen by A.D.Sakharov in 1967 \cite{Sakh1967}. Thus, while the Weyl tensor is gradually vanishing, the counter-term, linear in scalar curvature is becoming dominant. This is how the General relativity may come onto play. Note, that for this to happen, the scalar curvature must be non-zero. It means that the (effective) energy-momentum tensor is not already traceless, and we arrive at the situation with spontaneously broken conformal symmetry. Therefore, in such a scenario, the very process of particle creation plays the central role.

First, we found the general vacuum spherically symmetric solution with varying two-dimensional conformally transformed scalar curvature. This appears to be the one-parameter family (up to the overall conformal factor equal to the square of the radius). For the specific choice of the dependence of the radius on this scalar curvature it reduces to the famous Mannheim--Kazanas metrics \cite{ManKaz89}. Moreover, we showed that the Schwarzschild-(anti)de Sitter of General Relativity is also the member of out family. The latter indicates that the vacuum solution in the conformal gravity exhibits the effective cosmological constant.

Finally, we found two different types of the pure vacuum solutions, completely source-free. One of them has zero Weyl tensor, and it is conformally related to the homogeneous and isotropic cosmological models (but empty!). That one which is compact (with positive spatial curvature) we called ``the gravitational bubble'', it is quite suitable for the ``creation from nothing''. Another one, with non-zero Weyl tensor, is the best candidate for the vacuum fluctuations, the seeds for all the subsequent evolution.


\section{Generals}
\label{gensec}

The action integral of the conformal gravity should be invariant under the conformal transformation of the metric tensor $g_{\alpha\beta}=e^{2\omega}\tilde g_{\alpha\beta}$ (for the inverse metric tensor $g^{\alpha\beta}=e^{-2\omega}\tilde g^{\alpha\beta}$, $e^{2\omega}$ is called the conformal factor). It is well known that the so called Weyl tensor (a.k.a. the conformal tensor)  $C^{\alpha\phantom{0}\phantom{0}\phantom{0}}_{\phantom{0}\beta\gamma\delta}$ is conformally invariant ($C^{\alpha\phantom{0}\phantom{0}\phantom{0}}_{\phantom{0}\beta\gamma\delta}=\tilde C^{\alpha\phantom{0}\phantom{0}\phantom{0}}_{\phantom{0}\beta\gamma\delta}$). In addition, in four-dimensional space-time its square, $C^2= C_{\alpha\beta\gamma\delta}C^{\alpha\beta\gamma\delta}$, when multiplied by the square root of the metric determinant, $\sqrt{-g}$, is also conformally invariant. Thus, for in our four-dimensional space-time the only choice for the gravitational action integral is
\begin{equation}
S_{\rm grav}=-\alpha_0\int C^2\sqrt{-g}d^4x.
\end{equation}
The Weyl tensor $C^{\phantom{0}\phantom{0}\phantom{0}\phantom{0}}_{\alpha\beta\gamma\delta}$ is defined as the totally traceless part of the Riemann curvature tensor  $R^{\phantom{0}\phantom{0}\phantom{0}\phantom{0}}_{\alpha\beta\gamma\delta}$ obeying the same algebraic symmetries:
\begin{eqnarray}
&&C_{\alpha\beta\gamma\delta}=R_{\alpha\beta\gamma\delta}
-\frac{1}{n-2}(R_{\alpha\gamma}g_{\beta\delta}-R_{\alpha\delta}g_{\beta\gamma} \nonumber
 \\
&&-R_{\beta\gamma}g_{\alpha\delta}+R_{\beta\delta}g_{\alpha\gamma})
+\frac{R}{(n-1)(n-2)}(g_{\alpha\gamma}g_{\beta\delta}-g_{\alpha\delta}g_{\beta\gamma})
 \label{Weyltensor}
\end{eqnarray}

The total action is the sum of the gravitational and matter action integrals,
\begin{equation}
S=S_{\rm grav}+S_{\rm matter}=-\alpha_0\int C^2\sqrt{-g}d^4x+\int {\cal L}_m2\sqrt{-g}d^4x,
\end{equation}
and the matter energy-momentum tensor is defined, as usual, as
\begin{equation}
T_{\alpha\beta}=\frac{2}{\sqrt{-g}}\frac{\partial \sqrt{-g}{\cal L}_m}{\partial g^{\alpha\beta}}.
\end{equation}
The conformal invariance imposes some restrictions on the structure of $T_{\alpha\beta}$. First (and this is the commonly mentioned fact) the energy-momentum tensor should be traceless. Indeed, since the gravitational action is conformally invariant, so must be the matter counterpart. Consider the following conformal variations of the metric tensor; $\delta g_{\alpha\beta}=2e^{2\omega}\tilde g_{\alpha\beta}\delta\omega=2g_{\alpha\beta}\delta\omega$ ($\delta g^{\alpha\beta}=-2g^{\alpha\beta}\delta\omega$), then,
\begin{eqnarray}
0&=&\delta S_{\rm matter}=\delta\int {\cal L}_m2\sqrt{-g}d^4x
\equiv\frac{1}{2}\int T_{\alpha\beta} \sqrt{-g}\delta g^{\alpha\beta}d^4x  \nonumber
\\
&=&-\int T_{\alpha\beta} g^{\alpha\beta}\sqrt{-g}\delta \omega d^4x \Rightarrow {\rm Tr}\left(T_{\alpha\beta}\right)=0.
\end{eqnarray}
Second, let us compare two energy-momentum tensor for two metrics related by the conformal transformation, and consider another type of the metric variation that does not affect the conformal structure, namely, $\delta g^{\alpha\beta}=e^{-2\omega}\delta \tilde f^{\alpha\beta}$. We gave
\begin{eqnarray}
\delta S_{\rm matter}&=&\delta\int {\cal L}_m2\sqrt{-g}d^4x
\equiv\frac{1}{2}\int T_{\alpha\beta} \sqrt{-g}\delta g^{\alpha\beta}d^4x
\frac{1}{2}\int\tilde T_{\mu\nu} \sqrt{-\tilde g} \delta \tilde g^{\mu\nu}d^4x \nonumber
\\
& \Rightarrow& \sqrt{-g} T_{\alpha\beta} \delta g^{\alpha\beta}=\sqrt{-\tilde g} \tilde T_{\alpha\beta} \delta \tilde g^{\alpha\beta}.
\end{eqnarray}
Therefore,
\begin{equation}
T_{\alpha\beta}=e^{-2\omega}\tilde T_{\alpha\beta}, \quad
T^\alpha_\beta=g^{\alpha\gamma}T_{\beta\gamma}=e^{-4\omega}\tilde g^{\alpha\gamma} \tilde T_{\beta\gamma}=e^{-4\omega}\tilde T^\alpha_\beta.
\end{equation}
We see, that the conformally related metrics describe space-times with different (conformally related) contents.

In what follows, we will be interested in the spherically symmetric conformal gravity. The general spherically symmetric metrics can always (at least, locally) be written in the form
\begin{equation}
ds^2 = \gamma_{ik}dx^i dx^k-r^2(x)(d\theta^2+\sin^2\theta d\varphi^2).
\end{equation}
Here $\theta$ and $\phi$ are familiar spherical angels, so, $d\Omega^2=d\theta^2+\sin^2\theta d\varphi^2$ is the line element of the unit sphere, $r(x)$ is its radius (in that sense that the area of the sphere equals $4\pi r^2$), $x^i$ are the remaining two coordinates, the Latin indices $i,k=0,1$ (or $u,v$ in the case of the double-null coordinates). It is useful to make the so called $(2+2)$-decomposition for all the quantities that enter the equations. In our case of the conformal gravity it is convenient also to choose deliberately the conformal factor, this will simplify considerably all the subsequent calculations.

Let us write the line element as follows
\begin{eqnarray}
ds^2 &=& \gamma_{ik}dx^i dx^k-r^2(x)(d\theta^2+\sin^2\theta d\varphi^2) \nonumber \\
 &=& r^2(x)[\tilde\gamma_{ik}dx^i dx^k-(d\theta^2+\sin^2\theta d\varphi^2)]
 =r^2(x)[d\tilde s_2^2 -d\Omega^2],
\label{r2s2}
\end{eqnarray}
where $d\tilde s_2^2=\tilde\gamma_{ik}dx^i dx^k$. We denote the transformed entities by ``tilde'' (as was already done before), and the metrics $d\tilde s_2^2$ we will call ``truncated'' as well as all the corresponding quantities. Thus, the square of the radius was chosen as the conformal factor. Some of the details of this $(2+2)$-decomposition were presented in the paper\cite{we}. Here we will need only the final result for the square of the Weyl tensor,
\begin{equation}
\tilde C^2 \equiv \tilde C_{\alpha\beta\gamma\delta}\tilde C^{\alpha\beta\gamma\delta}=\frac{1}{3}(\tilde R-2)^2,
\end{equation}
where $\tilde R$ id the scalar curvature of the two-dimensional space-time described by the metrics $\tilde\gamma_{ik}$.


\section{Spherically Symmetric Bach Equation}
\label{SphBachsec}

The conformal gravity field equations obtained by varying the total action integral, are called the Bach equations. Here we will derive them already for the spherical case, by varying the truncated metrics $\tilde\gamma_{ik}$ only.
Let us start with the gravitational one;
\begin{eqnarray}
S_{\rm grav}&=&-\alpha_0\int C^2\sqrt{-g}d^4x=-\alpha_0\int \tilde C^2\sqrt{-\tilde \gamma}\sin\theta d^2xd\theta d\phi \nonumber \\
&=&-4\pi\alpha_0\int \tilde C^2\sqrt{-\tilde \gamma}d^2x =-\frac{4\pi}{3}\alpha_0\int \left(\tilde R -2\right)^2\sqrt{-\tilde \gamma}d^2x,
\end{eqnarray}
where we made use of the conformal invariance of the product $C^2\sqrt{-g}=\tilde C^2\sqrt{-\tilde g}=\tilde C^2\sqrt{-\tilde \gamma}\sin\theta$ and integrated over angle variables. Consider now the matter part of the action;
\begin{equation}
S_{\rm matter}=\int {\cal L}_m2\sqrt{-g}d^4x=4\pi\int r^4 {\cal L}_m2\sqrt{-\tilde \gamma}d^2x.
\end{equation}
The two-dimensional part of the original energy-momentum tensor equals
\begin{eqnarray}
T_{ik}&=&\frac{2}{\sqrt{-g}}\frac{\partial \sqrt{-g}{\cal L}_m}{\partial \gamma^{ik}}=
\frac{2\pi\partial (r^4 {\cal L}_m\sqrt{-\tilde\gamma})}{r^4\sqrt{-\tilde\gamma}\partial (\frac{1}{r^2}\tilde\gamma^{ik})}=
\frac{8\pi r^2\partial({\cal L}_m\sqrt{-\tilde\gamma})}{\sqrt{-\tilde\gamma}\partial (\tilde\gamma^{ik})}
\end{eqnarray}
since $r^2$ (the conformal factor) does not take part in the varying $\tilde\gamma^{ik}$. By definition (already used earlier), the transformed (``tilded'') energy-momentum tensor is
\begin{eqnarray}
\tilde T_{ik}&\equiv&\frac{8\pi}{\sqrt{-\tilde\gamma}}\frac{\partial (r^4 {\cal L}_m \sqrt{-\tilde\gamma})}{\partial \tilde\gamma^{ik}}=\frac{8\pi r^4}{\sqrt{-\tilde\gamma}}\frac{\partial {\cal L}_m \sqrt{-\tilde\gamma}}{\partial \tilde\gamma^{ik}}
\end{eqnarray}
thus,
\begin{equation}
T_{ik}=\frac{1}{r^2}\tilde T_{ik}, \quad T^i_k=\frac{1}{r^4}\tilde T^i_k.
\end{equation}
Let us start the procedure of the variation of the gravitational action integral.
\begin{eqnarray}
\delta S_{\rm grav}&=&-\frac{4\pi}{3}\alpha_0\int\left(2(\tilde R-2)\delta\tilde R\sqrt{-\tilde\gamma}+(\tilde R-2)^2\delta\sqrt{-\tilde\gamma}\right)d^2x \nonumber \\
&=&-\frac{8\pi\alpha_0}{3}\int\left(\frac{\tilde R^2-4}{4}\tilde\gamma_{ik}\delta\tilde\gamma^{ik}+(\tilde R-2)\tilde\gamma^{ik}\delta\tilde R^{ik}\right)\sqrt{-\tilde\gamma}d^2x.
\end{eqnarray}
Here we took into account that in two-dimensional space-time $\tilde R_{ik}=\tilde R\tilde\gamma_{ik}/2$. To calculate $\delta\tilde R^{ik}$ we will made use of the wonderful identities from the book by B.\,S.\;De\;Witt \cite{DeWitt}:
\begin{equation}
\delta R^{\alpha}_{\phantom{0}\beta\gamma\delta}=\left(\delta \Gamma^{\alpha}_{\beta\delta}\right)_{;\gamma}-\left(\delta \Gamma^{\alpha}_{\beta\gamma}\right)_{;\delta},
\end{equation}
\begin{equation}
\delta \Gamma^{\gamma}_{\alpha\beta}=\frac{1}{2}g^{\gamma\sigma}\left[\left(\delta g_{\sigma\alpha}\right)_{;\beta}+\left(\delta g_{\sigma\beta}\right)_{;\alpha}-\left(\delta g_{\alpha\beta}\right)_{;\sigma}\right].
\end{equation}
The semicolon denotes the covariant derivative with respect to the metric tensor $g_{\alpha\beta}$ (note, that $\delta \Gamma^{\alpha}_{\beta\gamma}$ is a tensor, and $(\delta g_{\sigma\alpha})_{;\beta}\neq0$).
In our case this gives
\begin{equation}
\delta \tilde R_{ik}=\left(\delta \tilde\Gamma^l_{ik}\right)_{|l}-\left(\delta \tilde\Gamma^l_{il}\right)_{|k},
\end{equation}
where the vertical line denotes the covariant derivative with respect to the two-dimensional truncated metrics $\tilde\gamma_{ik}$. Substituting all this into the variation of the action, integrating (several times) by parts and omitting total derivatives, we get, after lengthy  calculations,
\begin{equation}
\delta S_{\rm grav}=-\frac{8\pi\alpha_0}{3}\int\left(\frac{\tilde R^2-4}{4}\tilde\gamma_{ik}+\tilde R^{|p}_{|p}\tilde\gamma_{ik}-\tilde R_{|ik}\right)\delta\tilde\gamma^{ik}\sqrt{-\tilde\gamma}d^2x,
\end{equation}
\begin{equation}
\delta S_{\rm matter}=\frac{1}{2}\int \tilde T_{ik}\delta\tilde\gamma^{ik}\sqrt{-\tilde \gamma}d^2x.
\end{equation}
Thus, we arrive at the following Bach equations
\begin{equation}
\tilde B_{ik}=\frac{1}{32\pi\alpha_0}\tilde T_{ik},
\end{equation}
where $\tilde B_{ik}$ is the truncated two-dimensional Bach tensor,
\begin{equation}
\label{Bachtenstr}
\tilde B_{ik} = \frac{1}{6} \left[ \tilde R^{|p}_{|p} \tilde \gamma_{ik} - \tilde R_{|ik} + \frac{\tilde R^2 - 4}{4} \tilde \gamma_{ik}\right].
\end{equation}
The result, of course, coincides with that obtained in \cite{we} by traditional methods.

The important note: due to the tracelessness of the energy-momentum tensor (and, thus, of the Bach tensor as well), we do not need the separate equations for $B_2^2=B_3^3$, since (unlike in General Relativity) the they are the algebraic (not differential) consequences of the three two-dimensional equations written above.

In the end of this subsection we would like to present our field equations in yet another, vectorial, form, that may appear very useful in some applications. Let us introduce the invariant $\tilde \Delta$,
\begin{equation}
\tilde \Delta = \tilde \gamma^{ik} \tilde R_{, i} \tilde R_{,k}
\end{equation}
which is nothing more but the (Lorentzian) square of the normal to the curves $\tilde R=const$. When $\tilde \Delta<0$, these curves are time-like, and the truncated curvature $\tilde R$ can be used as a spatial coordinate. For $\tilde \Delta>0$, the curves are space-like, and $\tilde R$ can be used as a time-coordinate. Below we will show only the final result, the details of derivation can be found in \cite{we},
\begin{equation}
\label{vectorial}
\left( \tilde \Delta + \frac{1}{6} (\tilde R^3 - 12 \tilde R) \right)_{,i} = \frac{3}{8\pi\alpha_0} \left(\tilde T \tilde R_{,i} - \tilde T^k_i  \tilde R_{,k}\right),
\end{equation}
where $\tilde T=\gamma^{lp}\tilde T_{lp}$, and comma denotes the partial derivative. There are here only two equations of the three original ones. The remaining equation can be recovered by the use of the integrability condition together with the continuity equation for the energy-momentum tensor. Note, that if one of the coordinates id $\tilde R$, then
\begin{equation}
\tilde \Delta = \tilde \gamma^{\tilde R\tilde R}=-A, \quad
A=\frac{1}{6}(\tilde R^3 - 12 \tilde R)-\frac{3}{8\pi\alpha_0}\int \tilde T^\eta_\eta d\tilde R
\end{equation}
($\eta$ is the second coordinate).


\subsection{Vacuum solutions}

The vacuum solutions in the conformal gravity an be derived in two classes. One of them consists of the space-times with constant truncated curvature $\tilde R=\pm2$. We will consider them a little bit later. Now let $\tilde R\neq const$. Then, from our vectorial equation, it follows immediately that
\begin{equation}
\tilde\Delta=-\frac{1}{6}(\tilde R^3-12\tilde R+C_0),
\label{DeltaV}
\end{equation}
where $C_0$ is the integration constant. To proceed further we need to choose some definite coordinate system, let it be $\tilde R$ and the null coordinate $z$ (the line $z=const$ should be null). This can be called the Finkelstein-like coordinates. In this case the general form of the two-dimensional metrics is
\begin{equation}
ds_2^2=\tilde A d\tilde z^2+2\tilde B d\tilde zd\tilde R.
\label{ds2V}
\end{equation}
Therefore, the structure of the metric tensor $\tilde\gamma_{ik}$ and the inverse, $\tilde\gamma^{ik}$, is the following
\begin{equation}
\gamma_{ik}=\left( \begin{array}{cc}
\tilde A & \tilde B \\ \tilde B & 0
\end{array} \right), \quad
\gamma^{ik}=\left( \begin{array}{cc}
0 & \tilde B^{-1} \\  \tilde B^{-1} & -\tilde A\tilde B^{-2}
\end{array} \right).
\label{gammaV}
\end{equation}
and
\begin{equation}
\tilde\Delta=\tilde\gamma^{\tilde R\tilde R}=-\frac{\tilde A}{\tilde B^2}
=-\frac{1}{6}(\tilde R^3-12\tilde R+C_0).
\label{DeltaV2}
\end{equation}
Consider now the $(ik)=(\tilde R\tilde R)$ component of the vacuum Bach equations (it is the simplest one because $g_{rr}=0$). It reads $0=-R_{|\tilde R\tilde R}=\tilde\Gamma^{\tilde R}_{\tilde R\tilde R}$,
where $\tilde\Gamma^{\tilde R}_{\tilde R\tilde R}$ is the Christoffel symbol. It is easy to see that $\tilde\Gamma^{\tilde R}_{\tilde R\tilde R}=\tilde\gamma_{z\tilde R,\tilde R}=\tilde B_{,\tilde R}$. Thus the above equation tell us that $\tilde B=\tilde B(\tilde z)$, and it can be absorbed by the appropriate redefinition of null coordinate, $\tilde z\to z(\tilde z)$. So, the truncated metrics becomes
\begin{equation}
ds_2^2=A dz^2+2dzd\tilde R, \quad A=\frac{1}{6}(\tilde R^3-12\tilde R+C_0).
\label{ds2Vtrunc}
\end{equation}
If we would choose the second coordinate $\eta$ orthogonal to to $\tilde R$, then the diagonal form of the truncated metrics looks as follows
\begin{equation}
\label{2-dimfin}
d \tilde s^2_2 = A d \eta^2 - \frac{d \tilde R^2}{A} \, .
\end{equation}
The correspondence between these two forms of metrics is provided by the coordinate transformation
\begin{equation}
\label{transform}
z=\eta-\int\frac{d \tilde R}{A} \, .
\end{equation}
The metrics (\ref{2-dimfin}) resembles the famous vacuum spherically symmetric solutions in General Relativity (their two-dimensional parts), the truncated curvature $\tilde R$ playing the role of the radius $r$, and our solution is also static. At the same time, it must be remembered that the general form of the four-dimensional vacuum spherically symmetric space-time is, actually,
\begin{equation}
\label{4-dimfin}
d \tilde s^2_4 = \Phi^2 d \tilde s^2_2 + (d \theta^2+ \sin^2 \theta d \varphi^2) \, .
\end{equation}
The radius $r=\Phi$ is an arbitrary function of two variables.

There is the famous spherically symmetric static solution of the vacuum Bach equations, found by P.\;D.\;Mannheim and D.\;Kazanas \cite{ManKaz89}. We can easily recover such a solution with a special choice $r=\Phi(\eta, \tilde R)$. First of all, for statics we must require $r=\Phi(\tilde R)$, then transform our metrics to the coordinates $(\eta,r)$ and demand it should be of the form
\begin{equation}
\label{2-dimfin2}
d s^2_2 = r^2 d\tilde s_2^2=Fd\eta^2-\frac{dr^2}{F}.
\end{equation}
This is possible because of the still arbitrariness of $r=\Phi(\tilde R)$. Then,
\begin{equation}
\tilde R = -\frac{C_1}{r}+ C_2, \quad
F(r) = \frac{C_2}{2}-\frac{C_1}{6 r} + \frac{(4\!-\!C_2^2)}{2 C_1} r
+ \frac{C_2(C_2^2\!-\!12)\!-\!C_0}{6 C_1^2} r^2.
\end{equation}
Note the appearance of the two new constants.

\subsection{``Gravitational bubbles''}

Here we present the solutions with constant curvature, which we have called ``Gravitational bubbles''.
Let us introduce the double null coordinates:
\begin{equation}
 ds_2^2=\gamma_{ik}dx^i dx^k=2e^{2\omega}dudv=2Hdudv.
 \label{doublenull}
\end{equation}
The metric with a constant curvature $\tilde R$ in the double null coordinates $(u,v)$:
\begin{equation} \label{Rconst}
\tilde R=\frac{2}{H^3}(-HH_{,uv}+H_{,u}H_{,v})=const, \quad H=e^{2\omega},
\end{equation}
therefore
\begin{equation}
\omega_{,uv}=-\frac{\tilde R}{4}e^{2\omega}.
\end{equation}
This is the known Liouville equation in the differential geometry.

The two-dimensional metrics with constant curvature can be written in the form:
\begin{equation}
 ds_2^2=\frac{2}{k^2}\frac{dFdG}{\left(F-\frac{\tilde R}{4k^2}G\right)^2}
 =\frac{8}{|\tilde R|}\frac{d\tilde ud\tilde v}{(\tilde u\pm\tilde v)^2},
\end{equation}
where
\begin{equation}
F=\int e^{f(u)}du=\tilde u, \quad
G=\int e^{g(v)}dv=\left|\frac{2k}{\sqrt{\tilde R}}\right|\tilde v.
\end{equation}
The metric function $H$ depends on $\tilde u-\tilde v$ at $\tilde R>0$ and $\tilde u+\tilde v$ at $\tilde R<0$.
In the case $\tilde R>0$ we have $z=u-v$, $w_{uv}=-w_{zz}$. The Liouville equation takes the form
\begin{equation}
\omega_{,zz}=\frac{\tilde R}{4}e^{2\omega}
\end{equation}
with the solution
\begin{equation} \label{Rconst2}
z=\pm\int\frac{dy}{\sqrt{\tilde Re^{y}+c}}
=\mp2\int\frac{d\alpha}{\sqrt{\tilde R+c \alpha^2}}, \quad \alpha=e^{-y/2}.
\end{equation}
We see that, up to the overall conformal factor, the solutions with $\tilde R=const$ can be cast in the Robertson-Walker form. If $\tilde R = + 2$, the corresponding universes are empty and with zero Weyl tensor. This solution we called "the gravitational bubble".

\section{Conclusion and Discussions}

We considered the general structure of the spherically symmetric solutions in the Weyl conformal gravity. The Bach equation were derived and investigated for the special type of metrics, which can be considered as the representative of the rather general class. The complete set of the pure vacuum solutions is found, consisting of two distinctive classes. The first class of solutions contains the solutions with a constant two-dimensional curvature scalar, and the representatives are the famous Robertson--Walker metrics. One of them we have called the ``gravitational bubbles'', which is compact and has zero Weyl tensor. In result, we obtained the solution for the pure vacuum curved space-times (without any material sources) what is impossible in General Relativity.

This phenomenon makes is easier to create the universe from ``nothing''. There are two different spherically symmetric vacua (up to the conformal transformation). This is the manifestation of the spontaneous symmetry breaking, because the vacuum with $\tilde R=2$ is isotropic, while that one with $\tilde R=-2$ is not. This symmetry breaking is not what one wants and needs (emergent General Relativity, the possibility to have the matter energy-momentum tensor with nonzero trace and so on), but still this provides us with some hopes for a further exploration.

The second class of solutions is with a varying curvature scalar. We found its representative as the one-parameter family, which can be conformally covered by the thee-parameter Mannheim--Kazanas solution. The vectorial equation is constructed that reveals the same features of non-vacuum solutions.

Some feature of the vacuum solution with varying two-dimensional truncated scalar curvature $\tilde R$ outside the material source may explain the observed accelerated expansion of the universe. Indeed, let us assume that $\tilde R$ is an increasing function of the radius (as in General Relativity) and put the observer at the center. By adding the matter layer we will reduce more and more the constant $C_0$ in the solution, and this corresponds to the greater and greater values of the effective cosmological constant outside the layers. Eventually one comes to that the attraction of the test matter will be replaced by the repulsion. This is the pure kinematic  effect, depending on the position of the observer and the amount of matter between the observer and the observation point. Meantime, it does not mean that the universe as a whole is expanding with some acceleration.

\section*{Acknowledgments}

This study was funded by RFBR according to the research project 15-02-05038 a.


\begin{thebibliography}{99}


\bibitem{Weyl} H. Weyl, {\it Math. Zeit.} {\bf 2}, 384  (1918).

\bibitem{tHooft14} G. 't~Hooft, {\it Local Conformal Symmetry: the Missing Symmetry Component for Space and Time}, arXiv:1410.6675 [gr-qc].

\bibitem{ManKaz89} P. D. Mannheim  and D. Kazanas,   {\it Astrophys. J.} {\bf 342}, 635 (1989).

\bibitem{Edery98} A. Edery and M. B. Paranjape,  {\it Gen. Rel. Grav.} {\bf 31}, 1031 (1999); arXiv:astro-ph/9808345

\bibitem{Verbin09} Y. Brihaye and Y. Verbin, {\it Phys. Rev.} D{\bf 80}, 124048 (2009); arXiv:0907.1951 [gr-qc].

\bibitem{Vil82} A. Vilenkin,  {\it Phys. Lett.} {\bf B117}, 25 (1982).

\bibitem{Penr10} R. Penrose, {\it Cycles of Time: An Extraordinary New View of the Universe}, (The Random House, Londonб 2010).

\bibitem{Penr14} R. Penrose,  {\sl Found. Phys.} {\bf 44}, 557 (2014).

\bibitem{ZeldNovStar1974} Ya. B. Zel'dovich, I. D. Novikov and A. A. Starobinsky,   {\it ZhETF} {\bf 66}, 1897 (1974) [{\it JETP} {\bf 39}, 933 (1974)].

\bibitem{Zeld70} Ya. B. Zel'dovich,  {\it Pis'ma ZhETF} {\bf 12}, 443 (1970) [{\it JETP Lett.} {\bf 9}, 307 (1970)].

\bibitem{ZeldStar71} Ya. B. Zel'dovich, and A. A. Starobinskii,  {\it ZhETF} {\bf 61}, 2161 (1971) [{\it JETP} {\bf 34}, 1159 (1972)].

\bibitem{LukashStar74} Lukash V. N and A. A. Starobinskii,  {\it ZhETF} {\bf 66}, 1515 (1974) [{\it JETP} {\bf 39}, 742 (1974)].

\bibitem{Parker69} L. Parker,  {\it Phys. Rev.} {\bf 183}, 1057 (1969).

\bibitem{GribMam69} A. A. Grib  and S. G. Mamaev,   Yad. Phys. {\bf 10}, 1276 (1969) [{\it Sov. J. Nucl. Phys.} {\bf 10}, 722 (1970)].

\bibitem{ZeldPit71} Ya. B. Zel'dovich and L. P. Pitaevsky,    {\it Comm. Math. Phys.} {\bf 23}, 185 (1971).

\bibitem{Sakh1967} A. D. Sakharov,    {\it Sov. Phys. Doklady} {\bf 177}, 70 (1967).

\bibitem{LL2} L. D. Landau  and E. M. Lifsitz,  {\it Course of theoretical physics, vol 2. The classical theory of fields} (Forth edition, Butterworth--Heinemann, Oxford, 2000).

\bibitem{we} V. A. Berezin, V. I. Dokuchaev and Yu. N. Eroshenko, {\it Spherically symmetric conformal gravity and ``gravitational bubbles''}, arXiv:1412.2917 [gr-qc].

\bibitem{DeWitt} B. S. DeWitt,  {\sl Dynamical Theory of Groups and Fields} (Gordon and Breach, NY, 1965), Chapter 16.

\end{thebibliography}
\end{document}